\documentclass[a4paper]{jpconf}
\usepackage{graphicx}
\begin{document}
\title{Kalman Filter Tracking on Parallel Architectures}

\author{
Giuseppe Cerati$^1$,
Peter Elmer$^2$,
Steven Lantz$^3$,
Kevin McDermott$^4$,
Dan Riley$^4$,
Matev\v{z} Tadel$^1$,
Peter Wittich$^4$,
Frank W\"urthwein$^1$,
Avi Yagil$^1$
}
\address{$^1$ University of California - San Diego , La Jolla, CA, 92093, USA}
\address{$^2$ Department of Physics, Princeton University, Princeton, NJ 08540, USA}
\address{$^3$ Center for Advanced Computing, Cornell University, Ithaca NY 14853, USA}
\address{$^4$ Laboratory of Elementary Particle Physics, Cornell University, Ithaca NY 14853, USA}
\ead{cerati@cern.ch}

\begin{abstract}
Power density constraints are limiting the performance improvements of modern CPUs. 
To address this we have seen the introduction of lower-power, multi-core processors, but the future will be even more exciting. 
In order to stay within the power density limits but still obtain Moore's Law performance/price gains, it will be necessary to 
parallelize algorithms to exploit larger numbers of lightweight cores and specialized functions like large vector units.
Example technologies today include Intel's Xeon Phi and GPGPUs.
\\
Track finding and fitting is one of the most computationally challenging problems for event reconstruction in particle physics. 
At the High Luminosity LHC, for example, this will be by far the dominant problem. 
The need for greater parallelism has driven investigations of very different track finding techniques including Cellular Automata or 
returning to Hough Transform. The most common track finding techniques in use today are however those based on the Kalman Filter~\cite{Fruhwirth}. 
Significant experience has been accumulated with these techniques on real tracking detector systems, both in the trigger and offline. 
They are known to provide high physics performance, are robust and are exactly those being used today for the design of the tracking system for HL-LHC.
\\
Our previous investigations showed that, using optimized data structures, track fitting with Kalman Filter can achieve large speedup 
both with Intel Xeon and Xeon Phi. We report here our further progress towards an end-to-end track reconstruction algorithm 
fully exploiting vectorization and parallelization techniques in a realistic simulation setup.
\end{abstract}

\section{Introduction}

\begin{figure}[h]
\includegraphics[width=20pc]{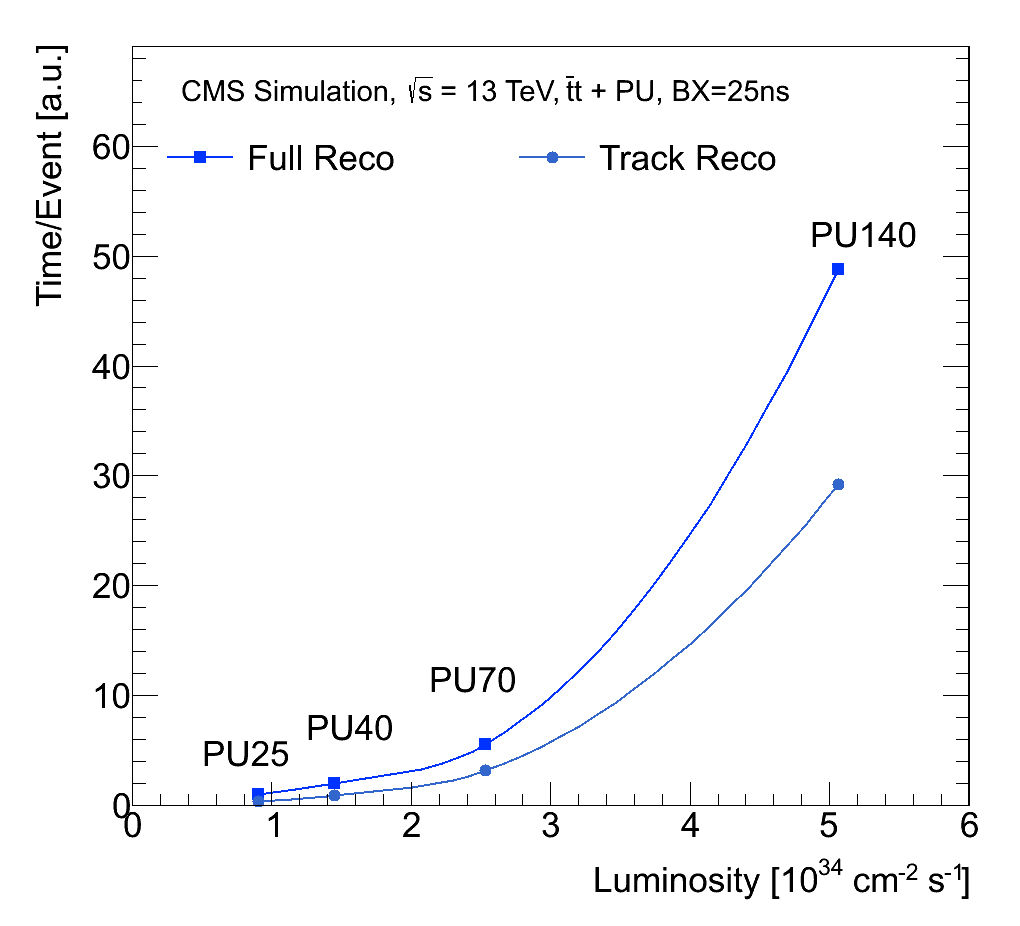}\hspace{2pc}%
\begin{minipage}[b]{14pc}\caption{\label{reco-time-pu} Timing vs luminosity for $t\bar{t}$+PU samples with 25 ns bunch crossing. 
Samples with different average pile-up are used as reported on the plot. 
Shown is timing of full CMS reconstruction and of the iterative tracking sequence~\cite{cerati}.}
\end{minipage}
\end{figure}

Pile-up (PU) represents a challenge for HEP event reconstruction, both in terms of physics performance and in terms of processing time. 
In fact, for PU values exceeding 100, as expected at the High Luminosity LHC (HL-LHC), the time needed for event reconstruction diverges 
(Fig.~\ref{reco-time-pu}); 
due to power density limitations to Moore's law, such a large increase is not sufficiently compensated by an increase in CPU clock frequency. 
For this reason, the reconstruction model currently used is most HEP experiments, based on traditional computers both for offline and online 
processing, cannot be sustained at HL-LHC without compromises between timing and physics performance that will impact the final sensitivity 
of the experiments.

As a solution to this problem we investigate a transition to highly parallel computing architectures such as e.g. 
Intel's Xeon Phi and NVIDIA GPGPUs.
The challenge in this context is that, due to distinct features of these architectures, 
a simple porting of the current, 'serial', implementation of the algorithms would be highly suboptimal:
code and data-structures need to be redesigned with a strong emphasis on hardware capabilities and limitations.

We initially focus our study on the Xeon Phi architecture. 
The main reason for this choice is that it shares many features with more conventional architectures, like the Intel Xeon, so that we can 
optimize and test the algorithm performance on both architectures at the same time;
however, there is no real prejudice on the choice of the architectures and in the future we plan to explore also other options, including GPGPUs. 
Xeon Phi coprocessors are characterized by 60 cores running up to 4 threads each and featuring 512-bit wide vector units;
thus, in order to achieve optimal performance all cores must be kept occupied, usage of vector units must be maximized, and code branching must be minimized.

We tackle the problem starting from the the most challenging algorithm, track reconstruction; 
tracking is by far the most time consuming process in event reconstruction (Fig.~\ref{reco-time-pu}) so that, 
if the proposed approach does not work in this case, there is little to gain from the rest of event reconstruction. 
Other algorithms, such as vertexing, jet clustering and so on will be studied at a later stage.
The present report describes the result of early studies and outlines the next steps of the project.

\section{Setup and Previous Results}

Below is a brief summary of the test setup and of previous results; a more detailed description can be found in~\cite{acat}.

In order to study the problem with maximum flexibility and with gradually increasing complexity, we developed a standalone tracking code
based on Kalman Filter and running on a simplified setup consisting of an ideal barrel geometry, a uniform and longitudinal magnetic field, 
gaussian-smeared hit positions, a particle gun simulation with flat transverse momentum distribution between 0.5 and 10 GeV, 
no material interaction and no correlation between particles nor decays.

The track reconstruction process can be divided in 3 steps: track seeding, building and fitting.
The track fit is the simple application of the Kalman Filter to a pre-determined set of hits, so it was a natural choice as a starting point.
It is implemented using Matriplex, a library for vectorized matrix operations we developed.
Matriplex is a matrix-major representation, where vector units elements are separately filled by the same element from different matrices; 
in other terms, $n$ matrices - and thus $n$ tracks - work in sync so that vectorization is used as an additional resource for parallelism.
The track fit using Matriplex gives same physics results and, even in serial case, it is faster than an equivalent version based on SMatrix. 

We tested the track fit both on Intel Xeon and Xeon Phi (native application) with OpenMP, obtaining similar qualitative results.
We observed a large speedup both from vectorization (4x/8x on Xeon/Xeon Phi) and parallelization (10x/80x).
The effective utilization efficiency of vector units is about 50\%; time scaling is close to ideal in case of 1 thread/core, while there 
is some overhead with 2 threads/core. 
Main performance limitations are related to data availability in L1 cache and data re-packing in Matriplex format.

As steps forward, we followed two distinct lines of development. 
On the one side we consolidated the track fitting results, identifying the performance bottlenecks to vectorization and progressing towards 
a fully realistic setup.
On the other side, we tackled the most time consuming part of the track reconstruction algorithm, redesigning the track building in order to
exploit vectorization and parallelization.

\section{Improvements to Track Fit}

\begin{figure}[h]
\includegraphics[width=20pc]{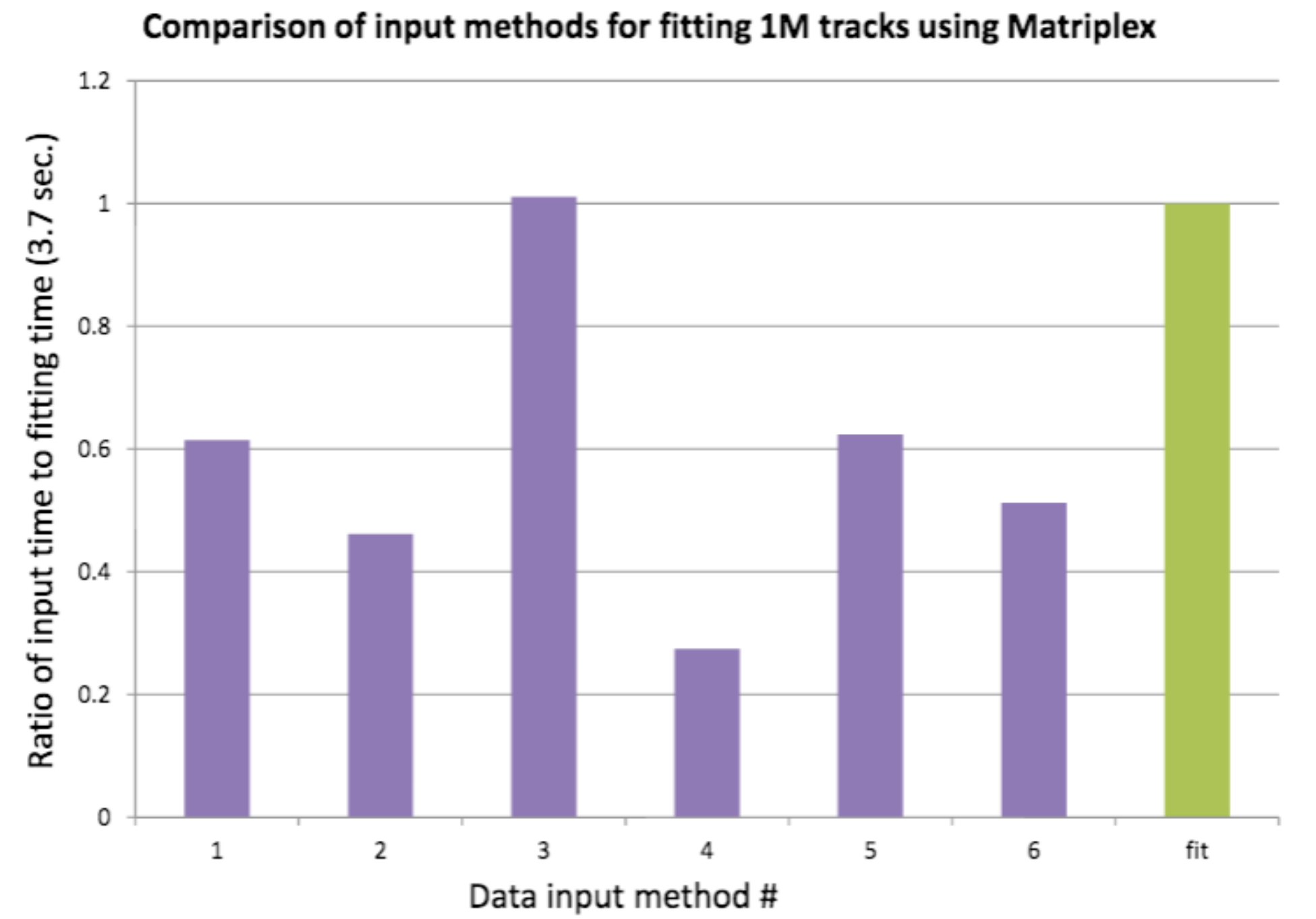}\hspace{2pc}%
\begin{minipage}[b]{14pc}\caption{\label{vectorize-data-input} \\
1. scatter data into Matriplexes, track by track, using simple for-loops \\
2. copy 16 tracks into contiguous memory, transpose into Matriplexes via loops \\
3. copy 16 tracks into memory, use Intel MKL to transpose each Mplex \\
4. copy 16 tracks into memory, vgather into Matriplexes by rows \\
5. copy 16 tracks into memory by rows, vscatter into Matriplexes \\
6. scatter data into Matriplexes, track by track, using Intel intrinsics
}
\end{minipage}
\end{figure}

A detailed profiling of the code showed that the relative fraction of time used for data input is large with respect to total fit time, 
where input refers both to data transfer to L1 cache and re-packing into Matriplex format.
Figure~\ref{vectorize-data-input} compares the time spent for data input using different methods to the actual fit time with Xeon Phi on 
a single-threaded execution using the vector units to the extent possible.
We explored different approaches (methods 1-3) and different intrinsics (4-6), 
either with a direct copy approach (1 and 6) or with a two-stage copy using temporaries (2-5), resulting in substantially different performance.
The best method (\#4) relies on two-stage copy and employs the vgather intrinsic.

\begin{figure}[h]
\begin{minipage}{17pc}
\includegraphics[width=17pc]{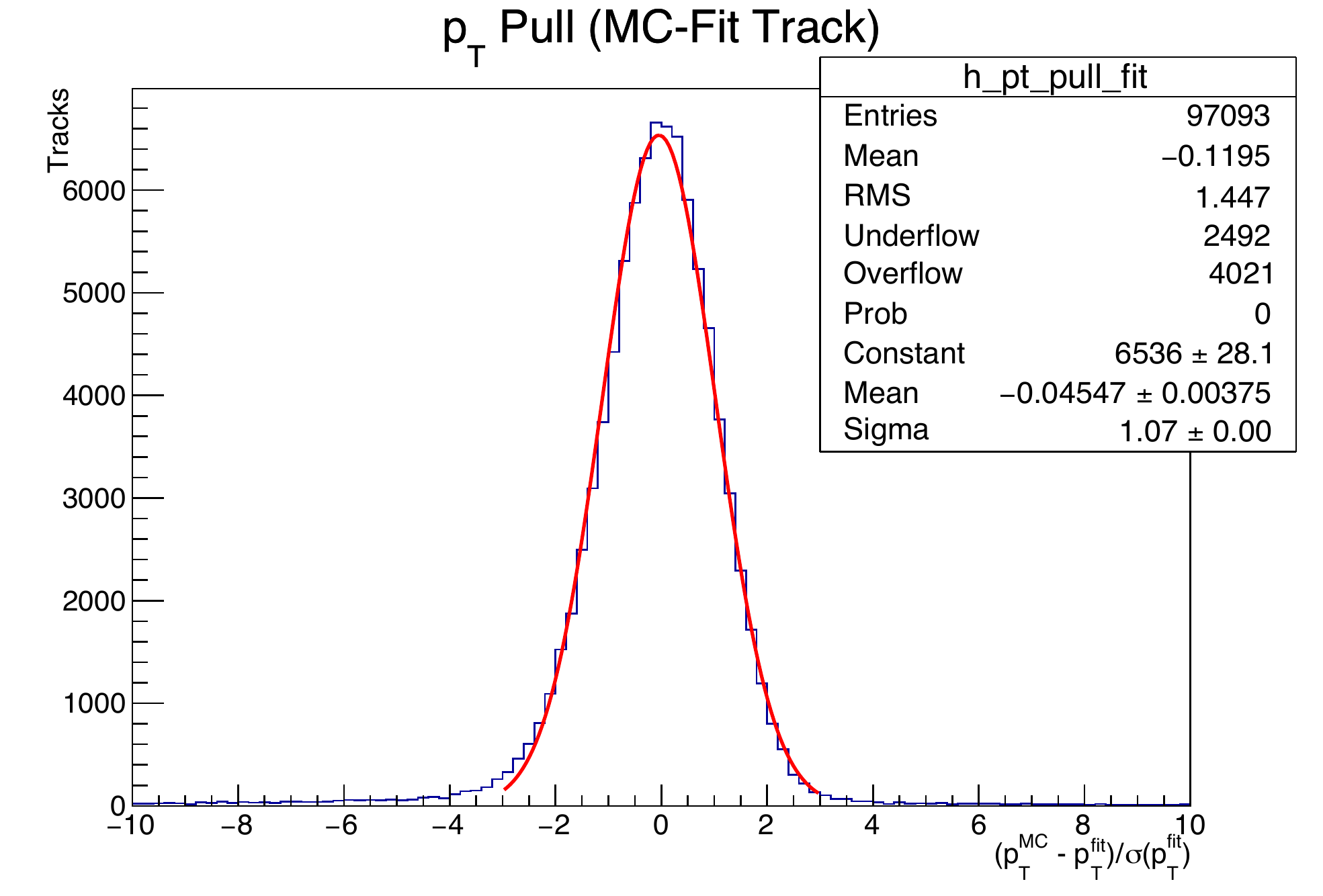}
\caption{\label{pTpull-cylindrical} Pull of the track transverse momentum using an barrel geometry composed of cylindrical layers.}
\end{minipage}\hspace{2pc}%
\begin{minipage}{17pc}
\includegraphics[width=17pc]{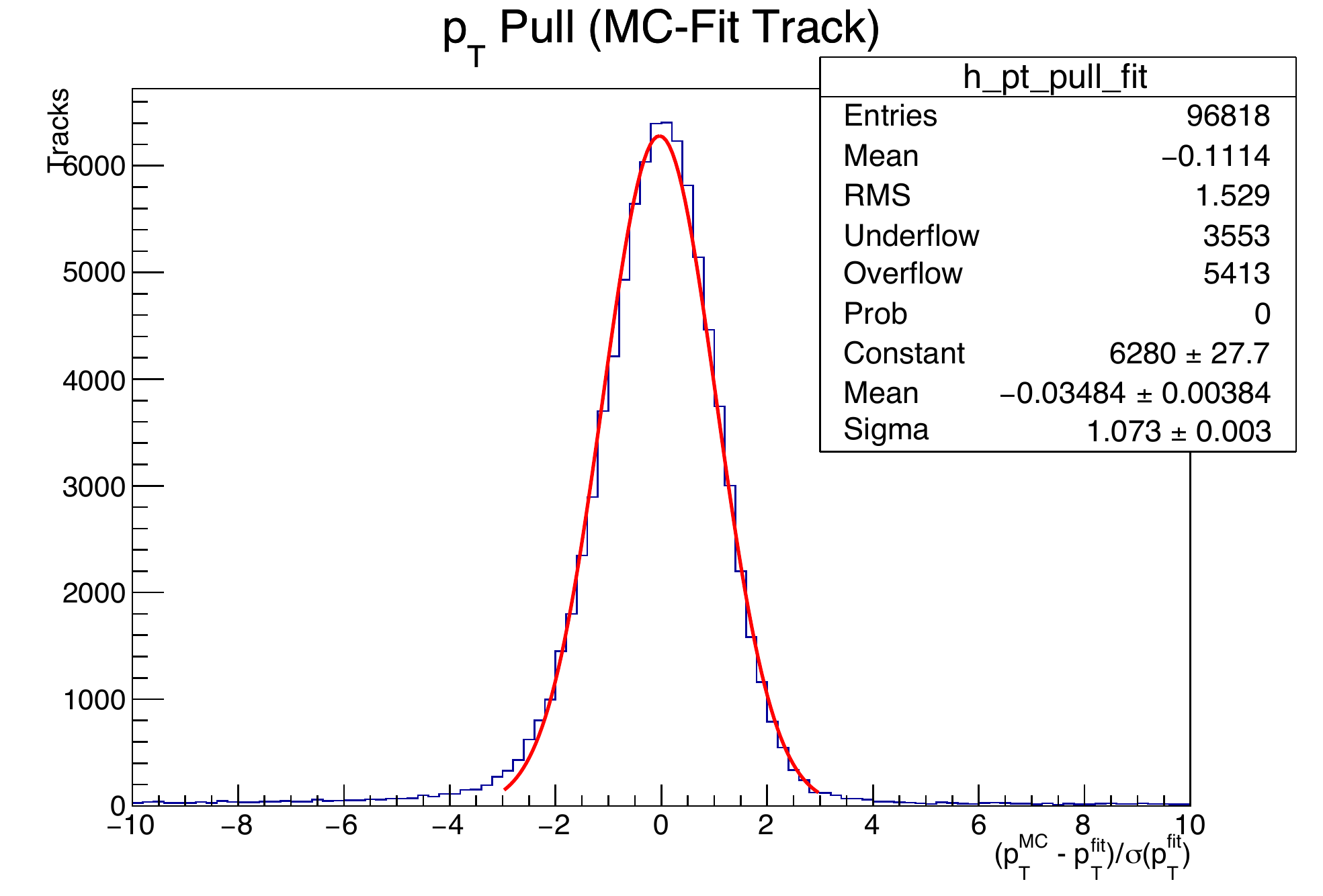}
\caption{\label{pTpull-polyhedric} Pull of the track transverse momentum using an barrel geometry composed of polyhedral layers.}
\end{minipage} 
\end{figure}

We also made significant progress towards the larger goal of implementing an end-to-end tracking chain in a realistic simulation, 
with track fitting as the final step. Our initially simple setup was designed to allow increases in complexity, so we can add more realistic 
features in an incremental way.  
In particular, we included a more realistic detector description implementing different barrel-like layouts by using the USolids geometry package~\cite{usolids}.
Such changes are transparent to the actual fitting code since the track parameters are propagated to the radius of the next hit in global coordinates.
We verified that the timing and physics performance of the track fit are not affected (Fig.~\ref{pTpull-cylindrical} and \ref{pTpull-polyhedric}).
Further improvements include the options to account for the effects of multiple scattering in the simulation and, 
instead of fitting the track hits as obtained from the simulation, to fit the tracks returned by the track building.

\section{Track Building}

The track building is the part of the track reconstruction process that, starting from a proto-track made of a 2-3 hits (the seed), 
identifies all other hits belonging to the track. 
From the computing point of view, it is by far to most expensive task in HEP event reconstruction.

While it entails similar calculations as the track fitting (i.e. track parameter propagation and update, $\chi^2$ of each measurement),
it adds two big complications to the problem. 
First, while during fitting the set of hits to process is already defined, in the building case a large number of hits is present at each layer, 
typically of the same order of the track multiplicity in the event.
Second, when more than one compatible hit is identified on the same layer, the combinatorial nature of the algorithm requires that a new track candidate 
is created for each compatible hit.

\begin{figure}[h]
\includegraphics[width=38pc]{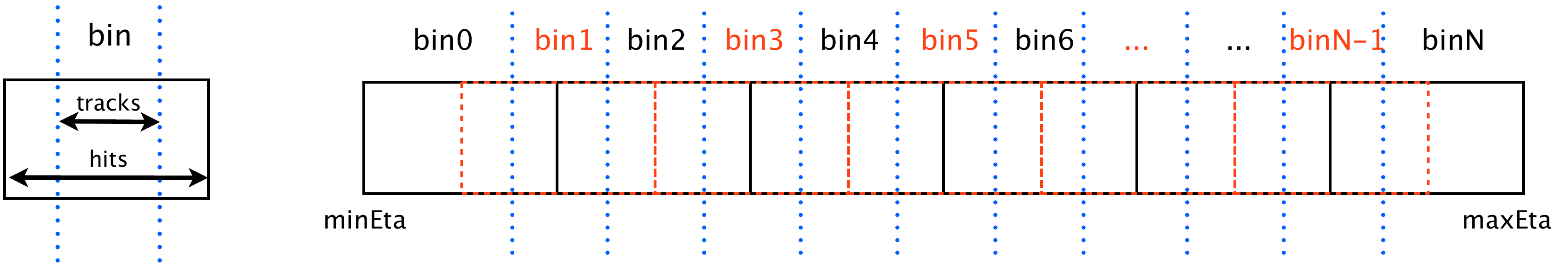}%\hspace{2pc}%
%\begin{minipage}[b]{20pc}
\caption{\label{eta-segmentation} Diagrams illustrating one single $\eta$ bin (left) and the full $\eta$ partitioning (right).}
%\end{minipage}
\end{figure}

Data locality is the key for reducing the large number of hits problem, so that only a reduced set of hits is considered at a given time and thus needs 
to be available in lower caches levels.
For this purpose, we partition the space both in pseudorapidity ($\eta$) and in azimuthal angle ($\phi$).
We exploit the fact that tracks do not bend in $\eta$ to define self-consistent $\eta$ partitions which are redundant in terms of hits,
so that there are no boundary effects and track candidates never search outside their $\eta$ bin (Fig.~\ref{eta-segmentation});
$\eta$ bins constitute a natural splitting for thread definitions.
A simple partitioning in $\phi$ is used for fast look-up of hits in the 3$\sigma$ compatibility window, where $\sigma$ is the uncertainty on the propagated
track $\phi$ on the considered layer.

\begin{figure}[h]
\begin{minipage}{17pc}
\includegraphics[width=17pc]{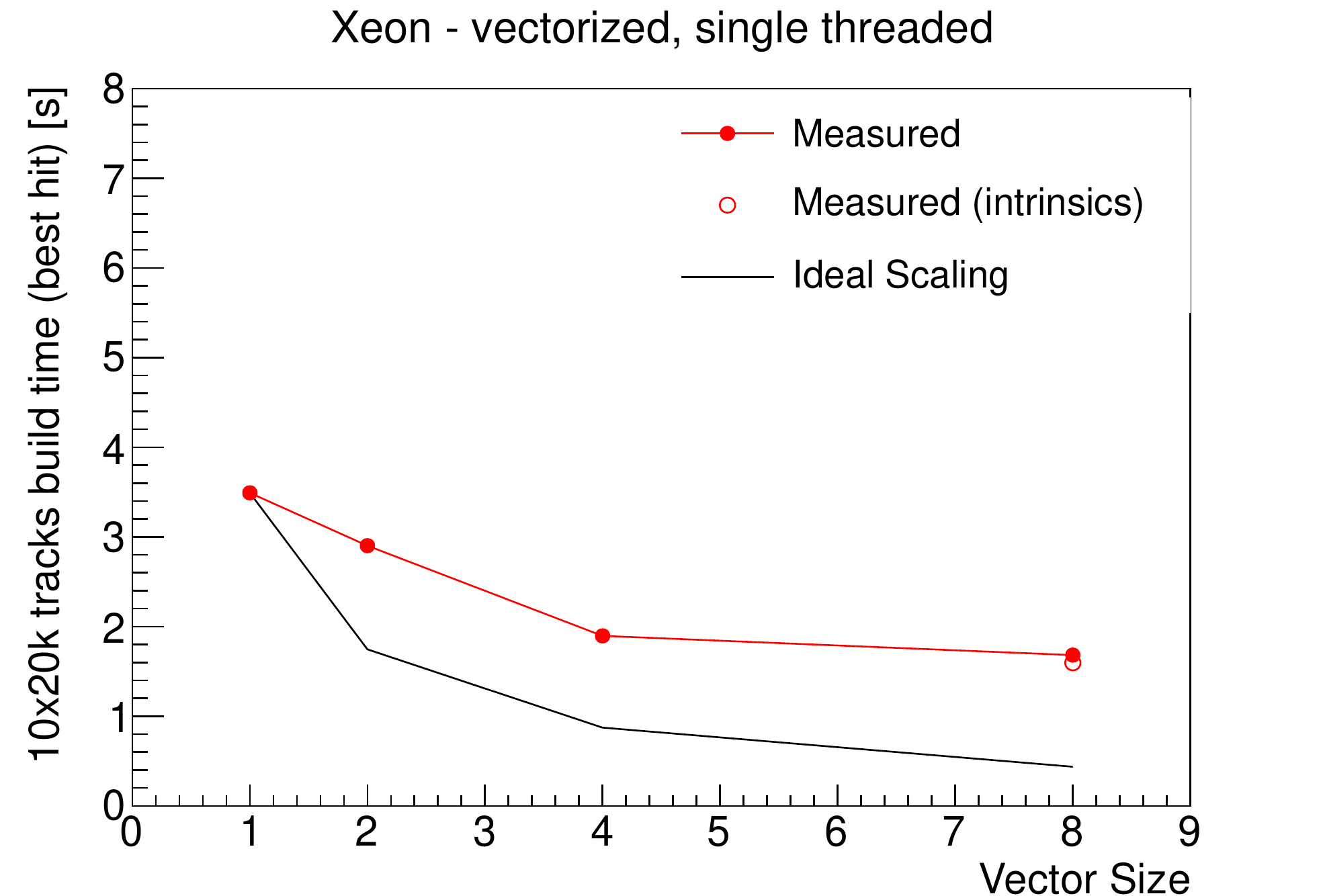}
\caption{\label{vector-host} Track building time (best hit version) vs number of elements in vector unit for Xeon processor.}
\end{minipage}\hspace{2pc}%
\begin{minipage}{17pc}
\includegraphics[width=17pc]{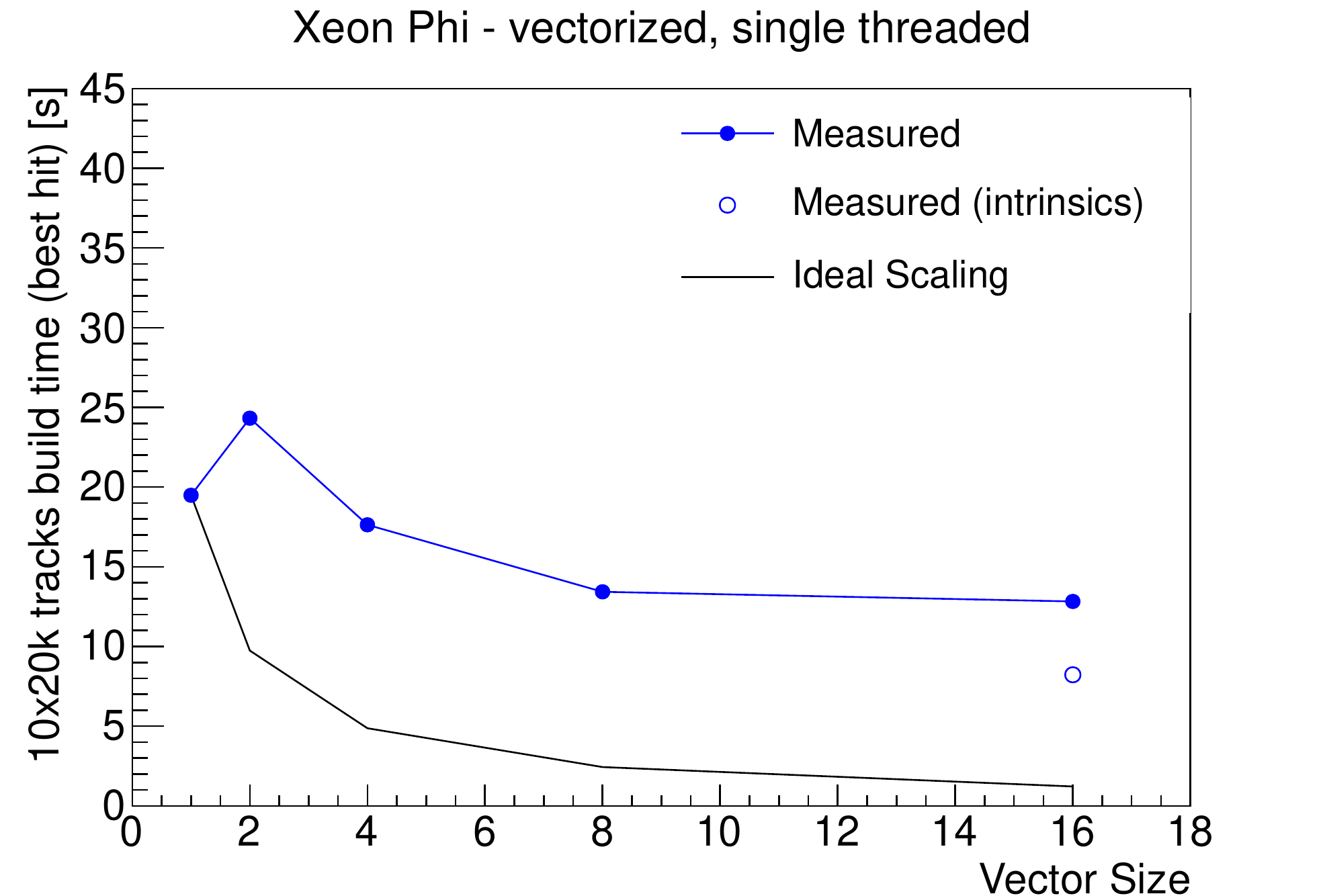}
\caption{\label{vector-mic} Track building time (best hit version) vs number of elements in vector unit for Xeon Phi.}
\end{minipage} 
\end{figure}

The two issues described above (i.e. hit multiplicity and branching of candidates) can be addressed separately by factorizing the problem in two stages.
In a first version of the algorithm, at each layer all hits within the compatibility window of a given track are probed and only the hit with the 
lowest $\chi^2$ is chosen, so that there is no multiplication of candidates; 
this simple and fast version already achieves good hit collection performance: 
running over 10 events with 20k tracks each, 70\% (93\%) of tracks are reconstructed with $\geq$90\% (60\%) of the hits.
In this setup we studied the performance in terms of vectorization, obtaining a maximum speedup of $>$2x both on Xeon 
and Xeon Phi (Fig.~\ref{vector-host} and \ref{vector-mic}). 
The scaling with vector unit size is monotonic on Xeon while an overhead is observed when vectorization is enabled on Xeon Phi, followed by a gradual speedup;
the usage of prefetching and gathering instrinsics gives a further time reduction when the vector units are maximally used.
It should be noted that, with respect to the fitting case, worse vectorization performance is expected since the latency due to data input
is amplified by the much larger number of hits processed at each layer.
%At this stage, the main limitation is that the set of hits to process on a given layer is not fully defined until the 
%propagation of the track parameters to the layer has completed; 
%threfore data copy and packing operations cannot be performed in advance causing executions stalls.

\begin{figure}[h]
\begin{minipage}{17pc}
\includegraphics[width=17pc]{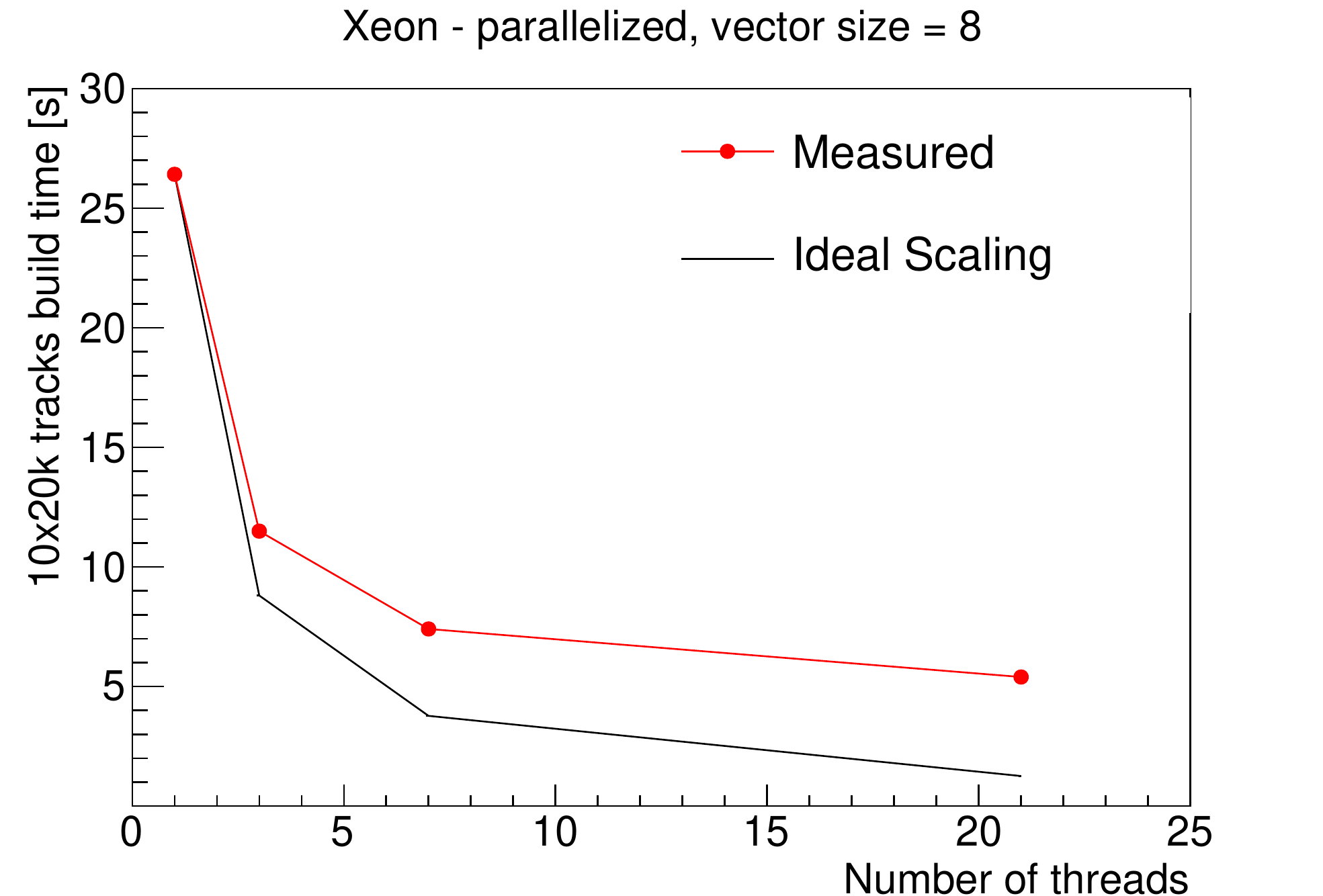}
\caption{\label{parallel-host} Track building time vs number of threads for Xeon processor.}
\end{minipage}\hspace{2pc}%
\begin{minipage}{17pc}
\includegraphics[width=17pc]{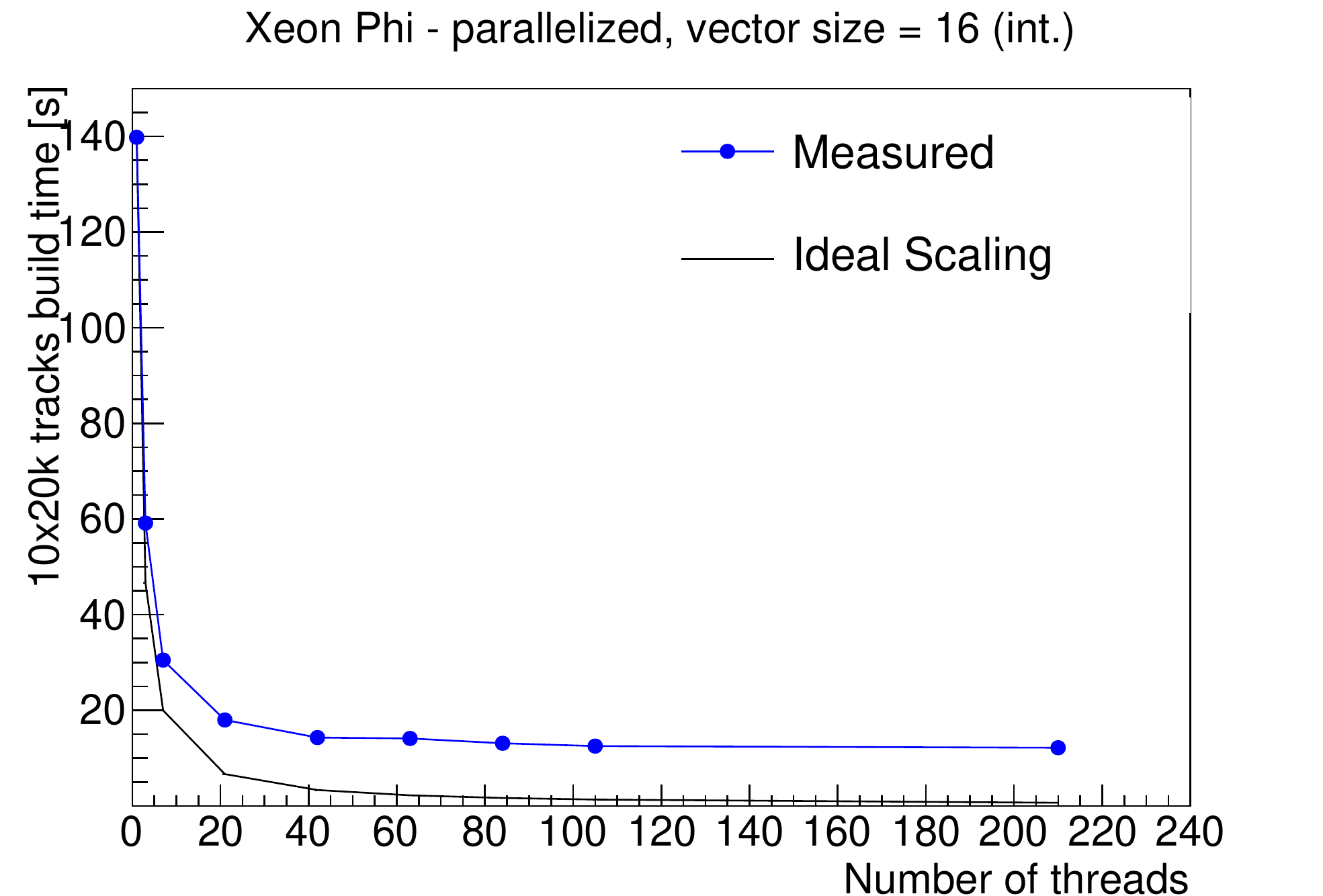}
\caption{\label{parallel-mic} Track building time vs number of threads for Xeon Phi.}
\end{minipage} 
\end{figure}

The second version is more advanced and targets ultimate physics performance. 
In this case we allow the branching of candidates, so that at each layer new candidates are created for all hits passing a $\chi^2$ cut of 30; 
also, in order to account for outlier hits and detector inefficiencies, candidates are allowed to miss the hit on one layer.
Up to 10 candidates per seed are considered and, when this number is exceeded, candidates are sorted based on the number of hits and the total $\chi2$ and 
those in excess are discarded.
Running over the same set of events of 20k tracks each, the hit collection performance is significantly improved and 85\% (95\%) of 
the tracks are reconstructed with $\geq$90\% (60\%) of the hits.
We tested the performance of this algorithm in terms of parallelization. 
Threads are evenly distributed threads across 21 eta bins, so that when the number of eta bins (nEtaBins) is a multiple of the number of threads (nThreads), 
n=nEtaBins/nThreads eta bins are processed in each thread, 
while for nThreads multiple of nEtaBins, seeds in each eta bin are distributed across n=nThreads/nEtaBins threads.
A speedup of 5x on Xeon and $>$10x on Xeon Phi is achieved (Fig.~\ref{parallel-host} and \ref{parallel-mic}); 
on Xeon Phi, we observe a saturation above nThreads=42 that will require further investigations.

\section{Conclusions and Outlook}

%my comments are “high level” and are mainly about the conclusions.
%It can be extended also to the intro, if you wish.
%I think that we want to remind in the summary that the setup is simplified and not realistic, as well as to maybe state the size of 
%the remaining challenge even for a simplistic case (i.e. that even in this case we are still far away from the gains we need). 
%Maybe list some of the next steps.
%You could mention in particular the challenge of dealing w branching during track building and its vectorization, the need to understand 
%what are the actual bottlenecks. whatever you feel like...
%Basically, make it clear that this is an ongoing r&d and not a solved problem.

In summary, a first implementation of a vectorized and parallelized track building in a simplified setup achieves significant speedup both on 
Xeon and Xeon Phi: 2x from vectorization on both architectures, 5x on Xeon and 10x on Xeon Phi from parallelization.
Track fitting studies have been consolidated by implementing an improved data input method and a more realistic detector geometry description.

The current results are definitely promising but many developments are still needed to deploy a complete tracking on parallel architectures.
The comparison with the scaling for ideal performance suggests a large margin for further improvements, so the first task is clearly to identify 
and address the current bottlenecks in the track building algorithm, both for vectorization and parallelization. 
In addition, the study of how to exploit vectorization in presence of branching points in the algorithm is far from being trivial and 
still needs to be addressed.
Lastly, the performance gain obtained on our simplified and standalone configuration will have to be propagated to an end-to-end track 
reconstruction sequence on a fully realistic setup.

%Additional developments could go in the direction of
%improving data locality by optimizing the partition size, by studying the performance as a function of the occupancy.
%performance studies at different track occupancy values;
%improve data locality: optimize partition size, smart sorting of candidates;
%complete transition to more advanced tools for parallelization (TBB).

\section{Acknowledgments}
This work was partially supported by the National Science Foundation, partially under Cooperative Agreement PHY-1120138, 
and by the U.S. Department of Energy.

\end{document}